\documentclass[
    ,final            
    ,sort&compress]  
  {aipproc}
\usepackage{amsmath}
\usepackage{subfig}
\layoutstyle{6x9}

\begin{document}

\newcommand{\citationneeded}{[!]}
\newcommand{\dee}{d}
\newcommand{\fig}[1]{Fig.~\ref{#1}}

\title{Higgs production via gluon fusion from $k_T$-factorisation}


\author{F. Hautmann}{
  address={Theoretical Physics Department, University of Oxford, Oxford OX1 3NP, UK}
}

\author{H. Jung}{
  address={Deutsches Elektronen Synchrotron, Hamburg 22603, Germany}
}

\author{V. Pandis}{
  address={Department of Applied Mathematics and Theoretical Physics, Cambridge, CB3 OWA, UK} 
}

\begin{abstract}
Theoretical studies of Higgs production via gluon fusion are
frequently carried out in the limit where the top quark mass 
is much larger than the Higgs mass, an approximation which reduces
the top quark loop to an effective vertex. We present 
a numerical analysis of the error thus introduced
by performing a  Monte Carlo calculation for $gg\rightarrow h$ in
$k_T$-factorisation, using the parton shower generator \verb|CASCADE|. 
We proceed to compare \verb|CASCADE| to the collinear
Monte Carlos \verb|PYTHIA|, \verb|MC@NLO| and \verb|POWHEG|. We study
the dependence of parton radiation on the resummation of
high-energy corrections taken into account by $k_T$-factorisation,
and its influence on predictions for the Higgs $p_T$ spectrum.
\end{abstract}

\maketitle

\section{Introduction}

In the $k_T$-factorisation method~\cite{ktfact1,ktfact2,ktfact3,ktfact4}  one
makes the transition from parton-level to hadron-level cross-sections 
through a convolution with the \emph{unintegrated} parton distribution 
functions~\cite{Hautmann07A,Hautmann07B}. By retaining the dependence on the transverse 
momenta and evaluating the cross-section with off-shell incoming partons, the
potentially large high-energy logarithms are automatically resummed.

Higgs boson production via gluon fusion is mediated through a top quark loop. In the \emph{heavy-top} limit in which 
$ 2m_t/m_H \gg 1$ this loop can be replaced by an effective vertex, reducing the loop count 
by one and simplifying the calculation~\cite{Nanopoulos78}. This approximation is frequently
used in theoretical studies of the Higgs so a quantification of the error 
introduced is important. Numerical
analyses in collinear factorisation indicate that the effects of the top-mass are
small when $2m_t/m_H < 1$~\cite{higlu,Pak09,Harlander09}. 

The cross-section of the top-quark triangle with off-shell initial-state gluons, having first been derived 
in the heavy-top limit~\cite{Hautmann02,Lipatov05}, now exists in the
literature with the full $m_t$ dependence~\cite{Pasechnik06,Marzani08}. It is therefore possible to examine the
impact of this approximation on both inclusive and exclusive quantities, now within
$k_T$-factorisation. A comparison of the cross-sections was carried out in Ref.\ \cite{Pasechnik06,Marzani09}
concluding that on the inclusive level corrections are of the order of 5\%.
Through the use of the $k_T$-factorised Monte Carlo \verb|CASCADE|~\cite{CASCADE1,CASCADE2} we 
confirm this finding and extend the investigation to the spectrum of the mini-jet radiation
accompanying the Higgs. 

An additional question, conceptually separate from the heavy-top approximation,
is the dependence of gluon radiation in association with Higgs production on the resummation of 
high energy corrections. We examine the impact of extra gluon radiation on the 
Higgs $p_T$ spectrum by comparing \verb|CASCADE| to the collinearly-factorised 
\verb|PYTHIA|~\cite{PYTHIA}, \verb|MC@NLO|~\cite{mcatnlo} and \verb|POWHEG|~\cite{powheg1,powheg2}.

All plots have been obtained for $pp$ collisions at $\sqrt{s}=14$~TeV and $m_H=120$~GeV. 

\section{Effects of the finite top-quark mass}
\begin{figure}
\centering
\begin{minipage}{2.5in}{\includegraphics[scale=0.25]{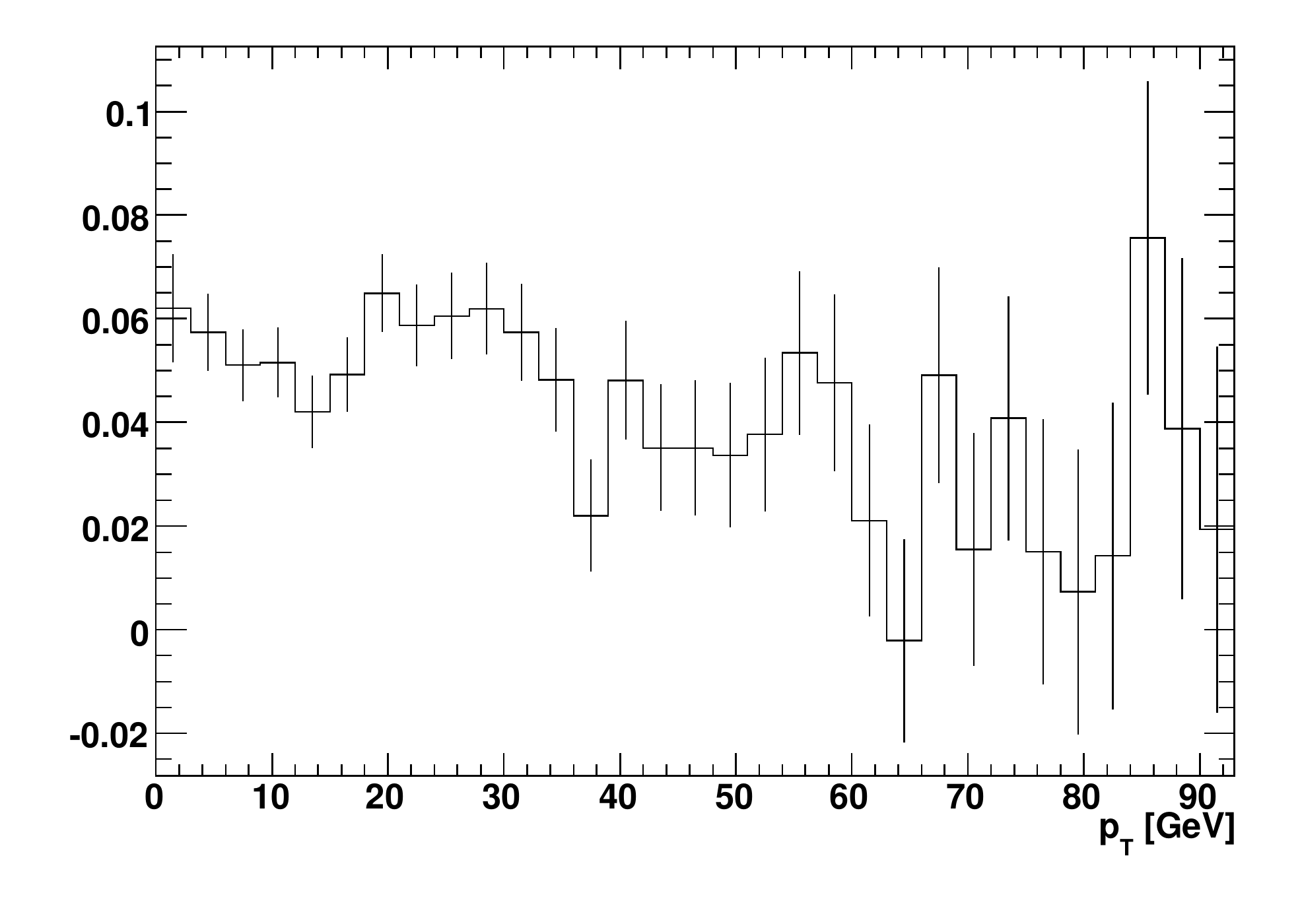}}\end{minipage}
\quad
\begin{minipage}{2.5in}\includegraphics[scale=0.25]{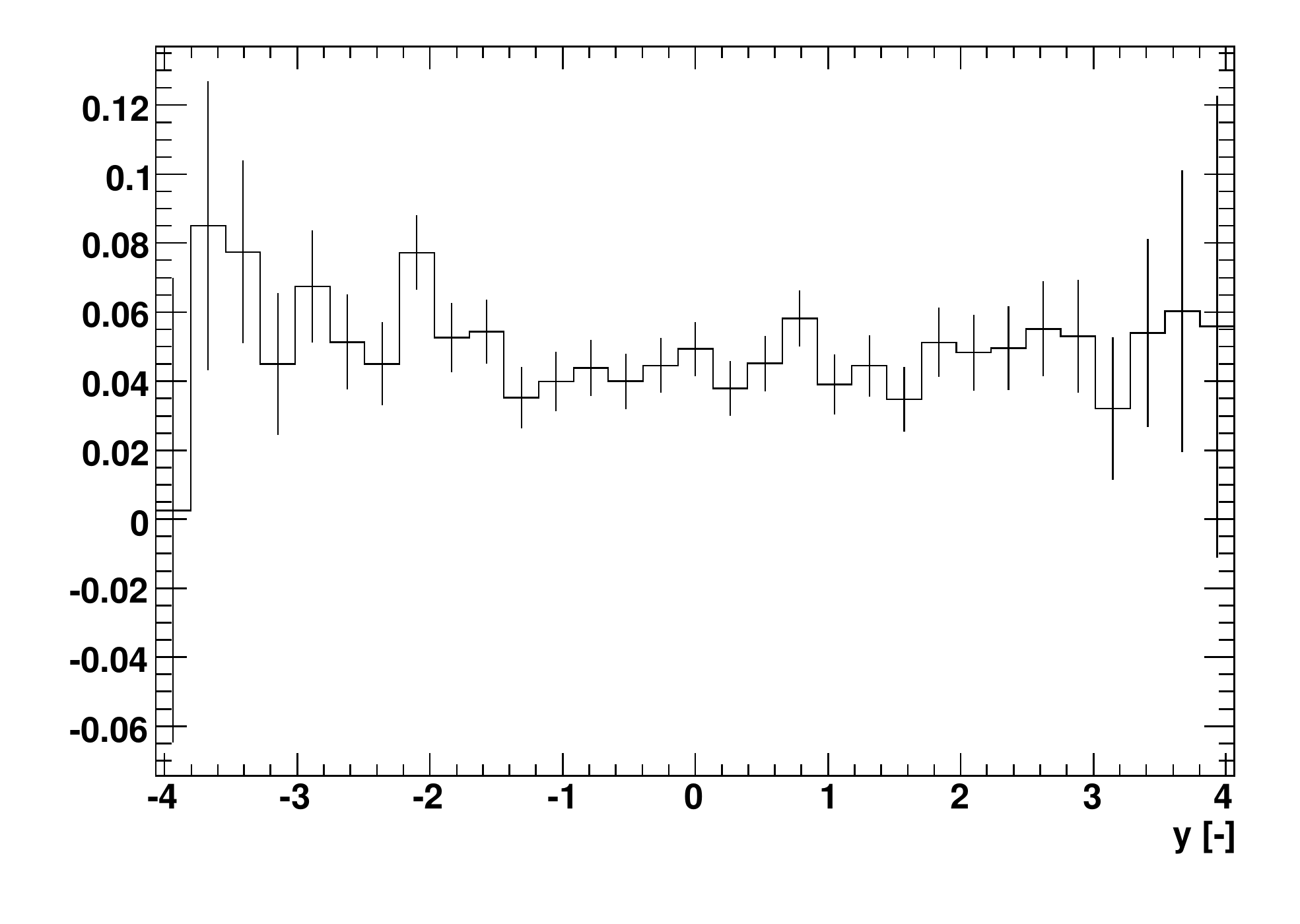} \end{minipage}
\caption{The quantity $1-\frac{\dee\sigma(m_\textrm{top}\rightarrow\infty)}{\dee W}/\frac{\dee\sigma(m_\textrm{top})}{\dee W}$ 
is plotted for $W=p^{\textrm{Higgs}}_T$ (left) and $W=y^{\textrm{Higgs}}$ (right), where y is the rapidity. The error bars reflect
only the statistical error of the Monte Carlo. These plots were obtained using the CCFM Set A~\cite{ccfmsetA,CASCADE1}.}
\label{fig:figs1}
\end{figure}

We find that retaining the full top-mass dependence in the matrix element leads to a
small and approximately uniform increase in the differential cross-sections 
in $p_T$ and $y$ of the order of 5\%. This is illustrated in Fig.~\ref{fig:figs1}
and is consistent with previous studies~\cite{Pasechnik06,Marzani09}.

\begin{figure}
\centering
\includegraphics[scale=0.4]{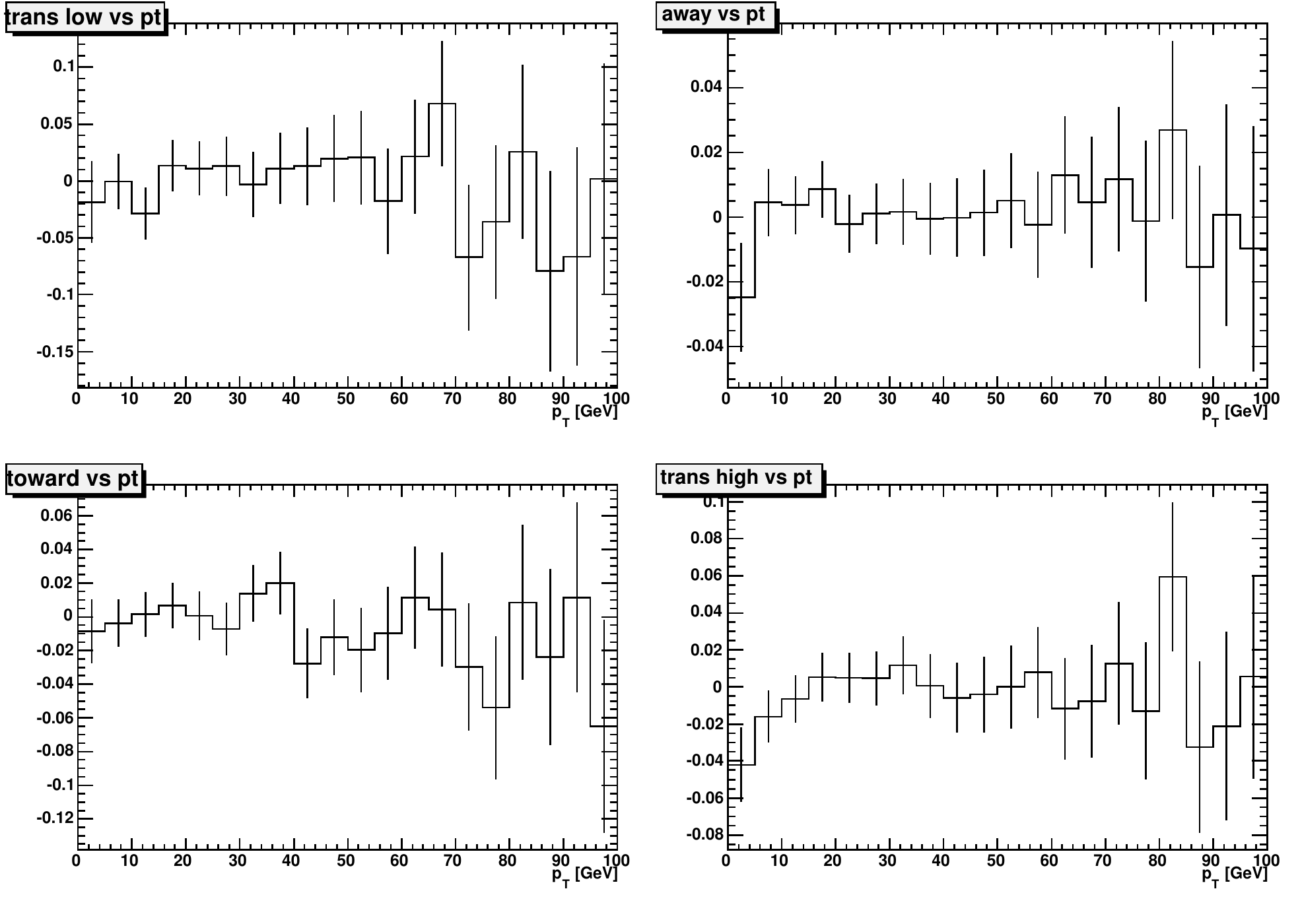}
\caption{The ratio $1-\frac{\dee N(m_\textrm{top}\rightarrow\infty)}{\dee p_T}/\frac{\dee N(m_\textrm{top})}{\dee p_t}$ 
is shown, where $N$ the number of mini-jets. The histograms are normalised to the $p_T$ spectrum of the Higgs and
thus do not scale with the cross-section.}
\label{fig:ratio-pt}
\end{figure}

We extend the study of top-mass effects by examining the mini-jet activity
accompanying Higgs boson production. We follow the underlying
event analyses of Ref.~\cite{Deak10,CDF02}, to which the reader is referred for the basic approach
and motivation. We divide the azimuthal plane
in four regions and accordingly classify the jets produced in association with
the Higgs. Jets are defined using the \verb|SISCone| algorithm~\cite{siscone} of the \verb|FastJet|~\cite{fastjet}
package with $R=0.4$ and $f=0.5$, applied on the hadron
level. We
impose the cut $p^{\textrm{jet}}_{T}>10$~GeV. The resulting multiplicity distributions in the 
four azimuthal regions are shown
in Fig.~\ref{fig:ratio-pt}, plotted against the Higgs transverse momentum.  They appear
to not be very sensitive to mass effects in the matrix element.

\section{Comparison to collinear Monte Carlos}
We compare the effect of resumming higher-order
contributions with a purely collinear description of
radiation. We extend the analysis of Ref.~\cite{Deak10} where 
\verb|CASCADE| was compared to the LO\footnote{PYTHIA also implements partial radiative higher-order corrections.} 
Monte Carlo \verb|PYTHIA| to include the collinear
NLO generators \verb|MC@NLO| and \verb|POWHEG|. 
Showering in \verb|MC@NLO| is performed through the angular-ordered \verb|HERWIG|~\cite{herwig1,herwig2} 
and we run \verb|POWHEG| coupled to the \verb|PYTHIA| shower. 
We operate \verb|PYTHIA| with the `new' underlying event model~\cite{Sjostrand04A,Sjostrand04B} (\verb|PYENVW|) 
using the `Perugia 0' tune~\cite{Skands10}.

\begin{figure}
\centering
\begin{minipage}{2.5in}{\includegraphics[scale=0.3]{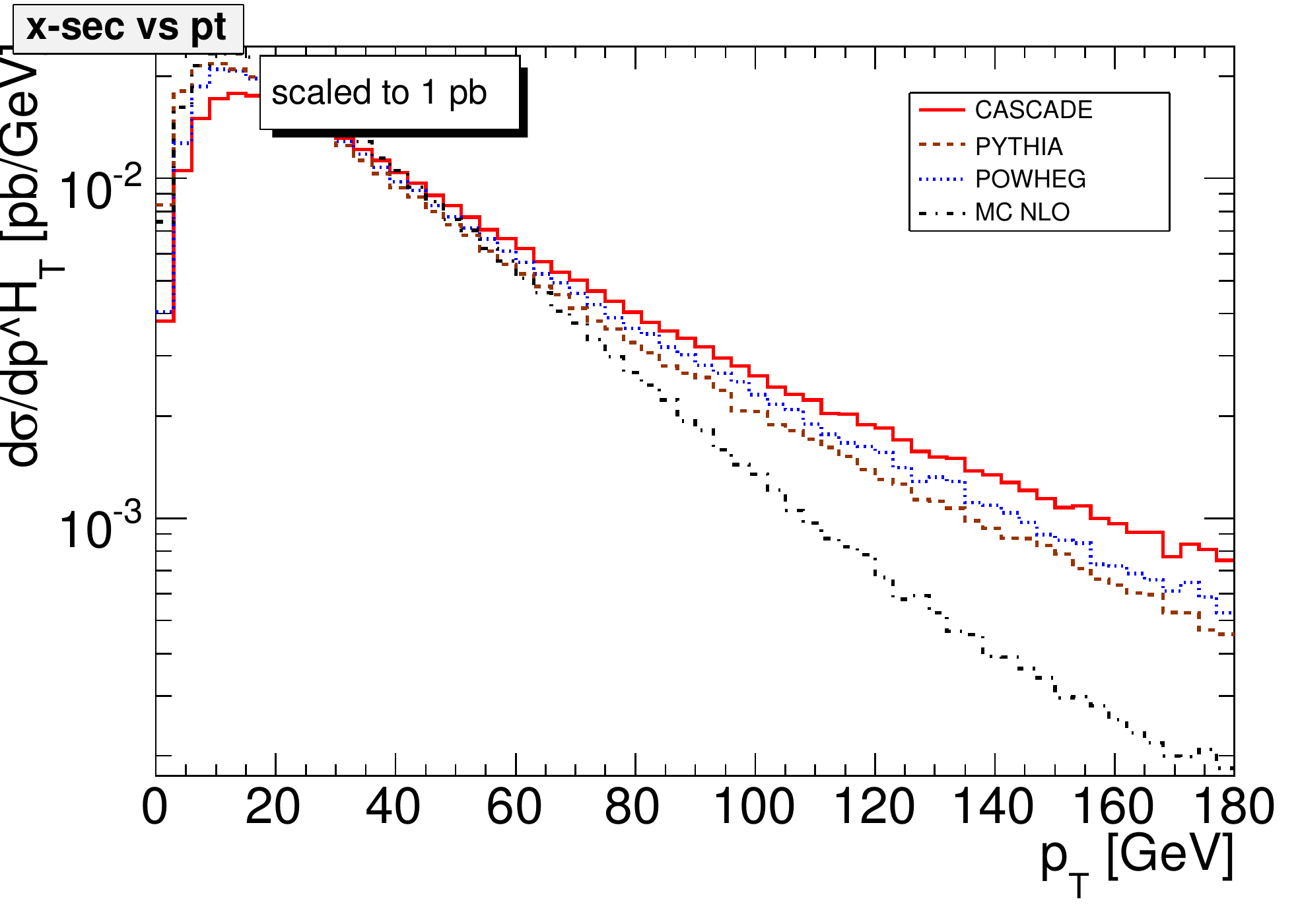}}\end{minipage}
\quad
\begin{minipage}{2.5in}\includegraphics[scale=0.3]{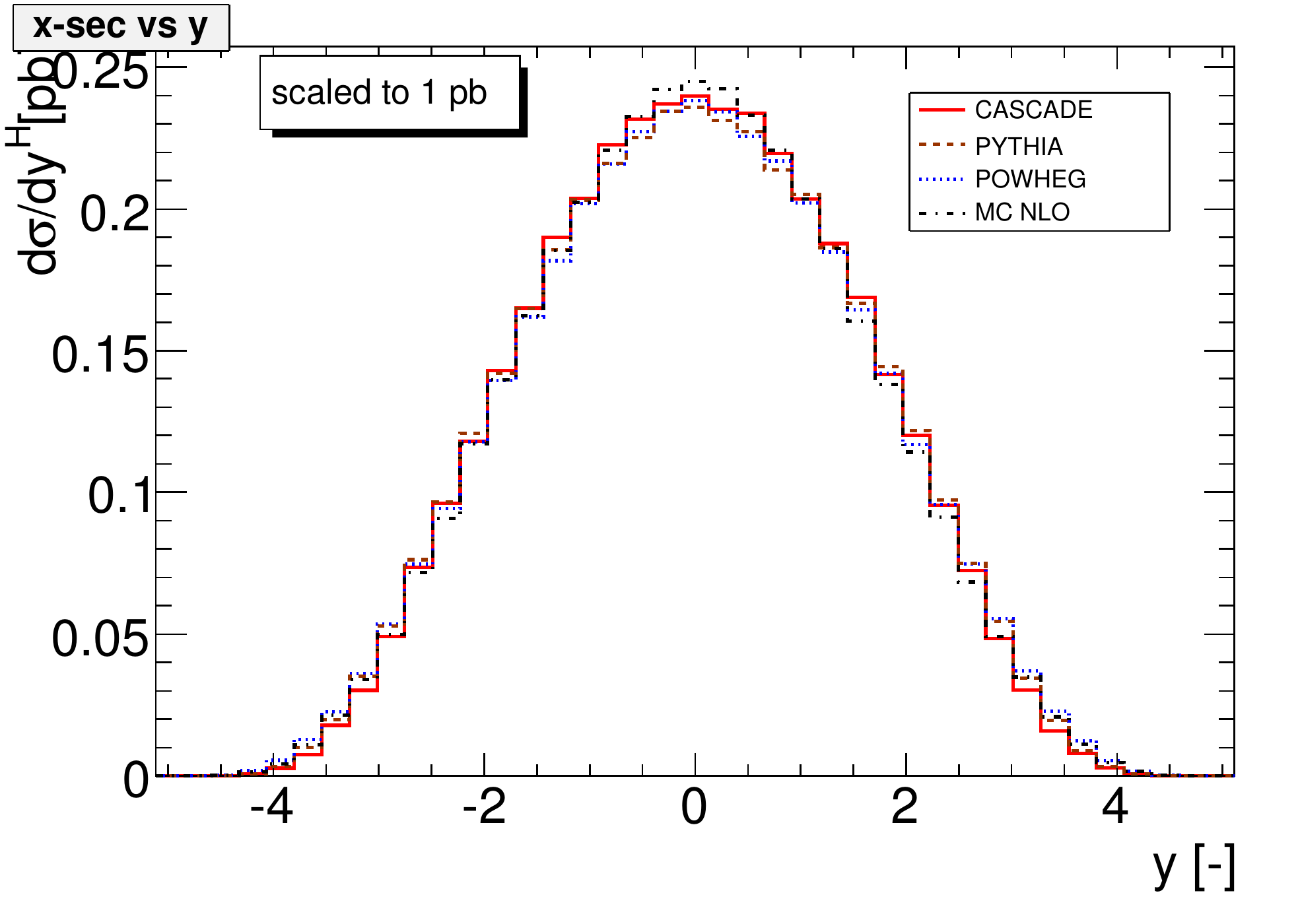} \end{minipage}
\caption{The $p_T$ (left) and rapidity (right) spectrum of the Higgs. The curves have been
scaled to a common token cross-section of $1$~pb.}
\label{fig:xsecs}
\end{figure}

In order to make a baseline comparison with the collinear
Monte Carlos we take the on-shell approximation for the hard matrix element 
$\textrm{ME}(\mathbf{k}_T) \rightarrow \textrm{ME}(\mathbf{k}_T = 0)\theta(\mu-k_T)$,
where $\mu$ is the factorisation scale. We use unintegrated gluon distributions obtained from 
deconvolution of the ordinary distributions as described in Ref.~\cite{ktfact1}. We
use one such standard set implemented in \verb|CASCADE|~\cite{CASCADE1,CASCADE2}. For the collinear generators we used the CTEQ6M~\cite{cteq6m} set.
The results are plotted in Fig.~\ref{fig:xsecs}.

Additional corrections to the matrix element
associated with the off-shellness contribute significantly to the spectrum.
The details of this will be elaborated on in a forthcoming publication. 
We find that the details of the initial-state showering are 
important even in the high-$p_T$ region.

\section{Conclusion}
We have implemented top-mass terms in the $k_T$-factorised
Monte Carlo \verb|CASCADE|. We have used this to analyse the uncertainty
induced by the heavy-top approximation that is commonly used to
simplify loop calculations. We have investigated this both for the
inclusive cross-section and the multiplicity of mini-jets accompanying
the Higgs boson.

Furthermore, we have examined the effect of the higher-order radiative terms
resummed by $k_T$-factorisation on the Higgs $p_T$ spectrum. We have
compared \verb|CASCADE| with the collinear Monte Carlos \verb|PYTHIA|, 
\verb|POWHEG| and \verb|MC@NLO|. We find that the impact of both the unintegrated gluon distributions
and matrix elements is significant even at $p_T$ of the order or higher
than the Higgs mass. 
%
\begin{theacknowledgments}
We would like to thank the organising committee of Diffraction 2010 for the invitation
to this splendid meeting. V.P.\ would like to thank the DESY directorate for their generosity and 
hospitality during his visit and also Mansfield College of the University
of Oxford for their financial assistance in his participation at this
conference.
\end{theacknowledgments}

\bibliographystyle{aipproc}   

\bibliography{proceedings}

\IfFileExists{\jobname.bbl}{}
 {\typeout{}
  \typeout{******************************************}
  \typeout{** Please run "bibtex \jobname" to optain}
  \typeout{** the bibliography and then re-run LaTeX}
  \typeout{** twice to fix the references!}
  \typeout{******************************************}
  \typeout{}
 }

\end{document}